\documentclass[a4paper,twoside]{article}

\usepackage{epsfig}
\usepackage{subcaption}
\usepackage{calc}
\usepackage{amssymb}
\usepackage{amstext}
\usepackage{amsmath}
\usepackage{amsthm}
\usepackage{multirow}
\usepackage{multicol}
\usepackage{pslatex}
\usepackage{apalike}
\usepackage{SCITEPRESS}     
\usepackage{bm}
\usepackage{booktabs}
\usepackage[linesnumbered,ruled,vlined,commentsnumbered]{algorithm2e}

\newcommand{\etal}{\textit{et al}.}
\newcommand{\eg}{\textit{e.g.}, }

\newcommand{\xvbox}[2]{\makebox[#1][l]{#2}}

\DeclareMathOperator*{\argmax}{arg\,max}

\begin{document}

\title{Human-Error-Potential Estimation\\ Based on Wearable Biometric Sensors}

\author{\authorname{Hiroki Ohashi\sup{1}\orcidAuthor{0000-0001-6970-2412} and Hiroto Nagayoshi\sup{1}}
\affiliation{\sup{1}Hitachi Ltd., R\&D group, Tokyo, Japan}
\email{\{hiroki.ohashi.uo, hiroto.nagayoshi.wy\}@hitachi.com}
}


\abstract{
This study tackles on a new problem of estimating human-error potential on a shop floor on the basis of wearable sensors.
Unlike existing studies that utilize biometric sensing technology to estimate people's internal state such as fatigue and mental stress, we attempt to estimate the human-error potential in a situation where a target person does not stay calm, which is much more difficult as sensor noise significantly increases.
We propose a novel formulation, in which the human-error-potential estimation problem is reduced to a classification problem, and introduce a new method that can be used for solving the classification problem even with noisy sensing data.
The key ideas are to model the process of calculating biometric indices probabilistically so that the prior knowledge on the biometric indices can be integrated, and to utilize the features that represent the movement of target persons in combination with biometric features.
The experimental analysis showed that our method effectively estimates the human-error potential.
}

\onecolumn \maketitle \normalsize \setcounter{footnote}{0} \vfill

\vfill
\section{\uppercase{Introduction}}
\label{sec:introduction}
Reducing human error is crucially important for almost all the industries to improve productivity, prevent defective products, and avoid serious accidents.
The development of IoT technology has advanced the analysis of 4M (Man, Machine, Material, and Method) factors, 
but it has been especially difficult to quantitatively analyze the factor of ``Man'' due to its uncertainty.
The uncertainty is attributed to various human factors including differences not only between workers but also within workers originating from the dynamically changing physical and mental states of individual workers.

This study aims to develop a method to estimate human-error potential of workers on a shop floor.
The visualization of human-error potential makes it possible to improve the working environment in various ways such as by suggesting a short break to workers who are found to have high error potential, by appropriately controlling air conditioning, and, more fundamentally, by reforming production lines in which the error potential of the workers tends to be higher.

There have been relevant studies that aim to utilize biometric sensing technology to estimate the internal state of humans such as fatigue~\cite{Sikander2019}, mental stress~\cite{panicker2019survey}, drowsiness~\cite{sahayadhas2012detecting}~\cite{ramzan2019survey}, and concentration~\cite{uema2017}.
The methods developed in these studies, however, cannot be trivially extended to estimate human-error potential on shop floors for two reasons.
First, there is no formal definition of human-error potential that can be directly used for formulating the estimation problem in a computationally tractable way.
Second, the previous studies implicitly assume that target persons are calmly seated, or at least they do not actively move, while workers usually move a lot on shop floors, resulting in producing undesirable noise on the measurement of the biometric sensors.

To overcome these difficulties, we first propose a novel formulation, in which the human-error-potential estimation problem is reduced to a classification problem.
We try to classify the workers' situation into different categories including a normal situation, where human-error-potential is expected to be low, and typical undesirable situations, where human errors have frequently occurred according to the literature.
Second, we propose a new method for estimating human-error potential in a situation in which target persons do not necessarily stay calm.
The key ideas are to model the process of calculating biometric indices probabilistically so that the prior knowledge on the biometric indices can be integrated to achieve robust estimation under noise, and to utilize the features that represent the movement of target persons in combination with biometric features.
The experimental analysis showed that our method effectively estimates the human-error potential.

The contributions of this study are summarized as follows.
\begin{enumerate}
    \item We propose, for the first time to the best of our knowledge, a formulation for estimating human-error potential on the basis of sensor data.
    \item We propose a new method for estimating human-error potential in a situation in which target persons do not necessarily stay calm.
    \item We experimentally verified the effectiveness of the proposed formulation and the method for estimating the human-error potential.
\end{enumerate}

\section{\uppercase{Related Work}}\label{sec:related_work}
There have been intensive efforts to categorize human errors from various aspects in order to systematically understand them and thereby prepare effective countermeasures to them.
Elwyn Edward developed the software, hardware, environment, liveware (SHELL) model, which helps to analyze the factors that are related to human error of workers in aviation systems, and Frank H. Hawkins later enhanced it and made it more widely accepted~\cite{Hawkins1987}.
Swain and Guttmann categorized the incorrect human outputs into two major classes: omission error and commission error~\cite{swain1983handbook}.
The latter is further divided into subcategories of selection error, error of sequence, time error, and qualitative error.
Rasmussen presented three levels of human behavior (skill-based, rule-based, and knowledge-based behavior) to make it possible to develop separate models for analysis~\cite{rasmussen1983skills}.
Another common method is to categorize errors into slips, lapse, and mistakes, which represent {\it ``actions-not-as-planned''}, {\it ``failures of memory''},
and {\it ``deficiencies or failures in the judgemental and/or inferential processes''},
respectively~\cite{reason1990}.
Although these studies are helpful for systematically analyzing past errors or preparing preventive measures in advance~\cite{hale1990human}\cite{edmondson2004learning}, they do not provide a concrete method for visualizing the live human-error potential on a shop floor, which can dynamically change.

The relevant studies from that viewpoint are those that have attempted to estimate people's internal state such as 
fatigue~\cite{Sikander2019}, mental stress~\cite{panicker2019survey}, drowsiness~\cite{sahayadhas2012detecting}~\cite{ramzan2019survey}, and concentration~\cite{uema2017}.
These studies used various biometric sensing technologies together with additional information.
Wijsman~\etal~\cite{wijsman2011towards} presented a method for detecting mental stress using electrocardiogram (ECG), respiration, skin conductance, and electromyogram (EMG).
Wang~\etal~\cite{wang2018novel} developed a method for detecting a driver's fatigue using electroencephalographic (EEG) signals.
Yamada and Kobayashi~\cite{yamada2018detecting} took a different approach for detecting  fatigue on the basis of eye-tracking data.
Tsujikawa~\etal~\cite{tsujikawal2018drowsiness} and Sun~\etal~\cite{sun2018neural} used video data for estimating the drowsiness.
Uema and Inoue~\cite{uema2017} developed a glasses-type sensor for estimating the level of concentration on the basis of electrooculography (EOG).
These studies, however, cannot be trivially extended to estimate human-error potential on shop floors mainly because they implicitly assume that target persons are calmly seated or at least do not actively move, whereas workers usually move a lot on shop floors, resulting in producing undesirable noise in the measurement of the biometric sensors.
Sun~\etal~\cite{sun2010activity} proposed an activity-aware mental stress detection method by using an accelerometer in combination with ECG and galvanic skin response (GSR) sensors, but the activities involved in their study were limited to rather simple ones, namely, sitting, standing, and walking.

The present study proposes a way to formulate the human-error-potential estimation as a classification problem, standing upon the characteristics of the human-error potential that have been revealed in the above-mentioned works.
Then we propose a new method for estimating human-error potential in a situation in which target persons work on a pseudo industrial operation.

\section{\uppercase{Formulation of human-error-potential estimation}}
\subsection{Formulation Based on Major Cause of Human Errors}
As described in Hawkins's SHELL model~\cite{Hawkins1987} and Swain and Guttmann's performance shaping factor (PSF)~\cite{swain1983handbook}, human-error is caused by various factors including not only the worker in question, but also other workers, environment, software, and hardware.
Ideally, all these factors should be sensed and taken into account when estimating human-error potential, but this study focuses only on the worker in question, which is naturally deemed most important, as the first step and leaves the other factors for future works.
To make the model as general as possible without relying on a specific task's characteristics, we focus on the possible root cause attributed to general psychological characteristics rather than focusing on the resultant categorization such as slip-lapse-mistake~\cite{reason1990}, and omission-commission~\cite{swain1983handbook}.
In terms of the skill-rule-knowledge (SRK) model by Rasmussen~\cite{rasmussen1983skills}, we focus on the factor of ``skill'' since rule and knowledge are not expected to dynamically change on shop floors and therefore do not have to be sensed in a live manner.
Among the skill-based errors, the statistics by Williamson~\etal~\cite{williamson1993human}, who studied 2000 incident reports, showed that ``inattention'' and ``haste'' were the major contributing factors.
Inattention was found to be closely related to multi-tasking by Ralph~\etal~\cite{ralph2014media}.

Standing upon these research outcomes, we attempt to build a model that can detect the situations where human-error tends to occur more frequently according to the statistics.
In this view, we can reduce the human-error-potential estimation problem to a classification problem of a worker's situation into a normal situation and the situation where human-error potential is deemed higher than usual.
More formally, we formulate the problem as the classification of three situations: a worker is working normally, in a hurry, and has to multi-task.
We call the three conditions the normal condition, time-pressure condition, and multi-task condition, respectively.

\subsection{Task Setting}
\begin{figure}
	\centering
	\includegraphics[width=1.0\linewidth]{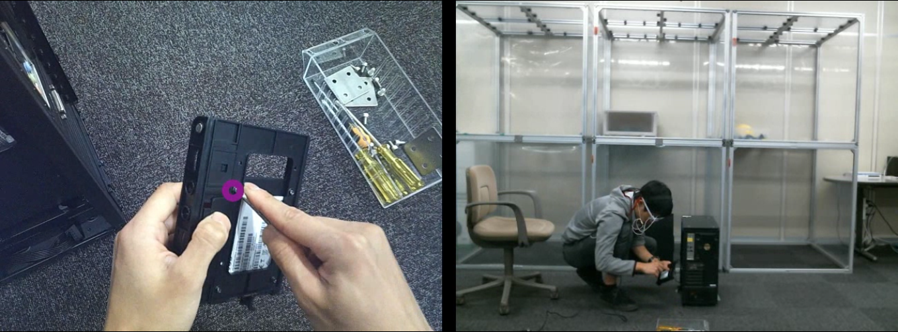}
	\caption{The SSD replacement task. Left: the video from eye-tracking glasses, where the purple circle denotes the estimated gaze location. Right: the video from a fixed camera.}
	\label{fig:exp_overview}
\end{figure}

We selected the replacement of a desktop PC's SSD as the experimental task because it contains basic operations that are commonly done in a wide variety of shop floors, \eg screw tightening, wiring, and assembly.
The specific procedure of the task is as follows: 
remove the screws of the side cover of the PC by hand,
remove the side cover,
pull out the SSD mount,
unplug the cables attached to the SSD,
remove the screws using a screwdriver to detach the SSD from the mount,
replace the SSD with a new one,
install the screws to attach the new SSD to the mount using the screwdriver,
plug the cables into the SSD,
place the SSD mount back to the original position,
and install the screws to attach the side cover (see Figure \ref{fig:exp_overview}).
Hereinafter, we call one series of this procedure a ``{\it trial}''.
Trials lasted three minutes on average, but they differed depending on the conditions and subjects.

In the time-pressure condition, we and asked the subjects to finish the trial within a defined time limit.
To increase the pressure, we told them the elapsed time every 30 seconds and told them every 10 seconds in the last 30 seconds.
In the multi-task condition, we gave them mental arithmetic problems of two digit addition and subtraction.
We read out the problems and the subjects had to answer out loud while working on the SSD-replacement task.
We did not give a time limit for these math questions, or any penalty for wrong answers.

\subsection{Sensor Selection}\label{sec:sensor}
\begin{figure}[t]
	\begin{minipage}{\linewidth}
    	\begin{minipage}{0.52\linewidth}
    		\centering
    		\includegraphics[width=0.99\linewidth]{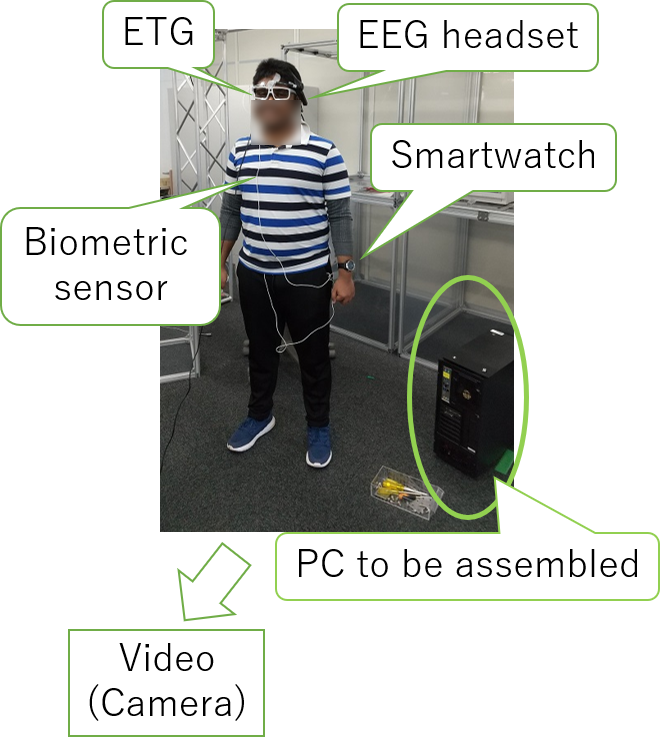}
    		\subcaption{Subject with all sensors.}
    	\end{minipage}\hfill
    	\hspace{3pt}
    	\begin{minipage}{0.45\linewidth}
        	\begin{minipage}{\linewidth}
        		\centering
        		\includegraphics[width=0.99\linewidth]{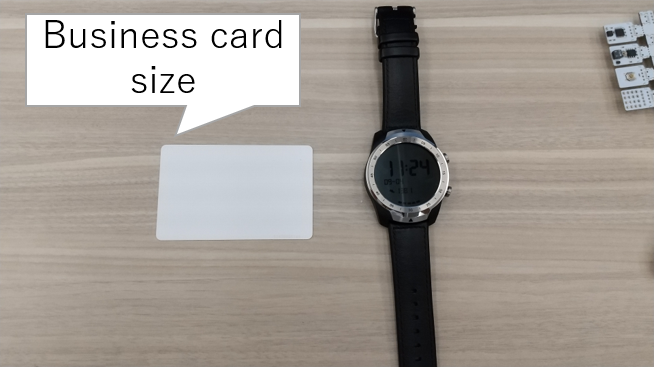}
        		\subcaption{Smartwatch.}
        		\vspace{3pt}
        		\centering
        		\includegraphics[width=0.99\linewidth]{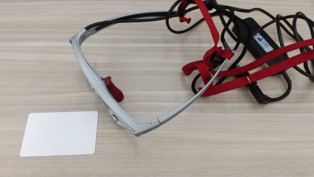}
        		\subcaption{Eye-tracking glasses (ETG).}
        	\end{minipage}\vfill
    	\end{minipage}\hfill
		\vspace{5pt}
    	\begin{minipage}{0.45\linewidth}
    		\centering
    		\includegraphics[width=0.99\linewidth]{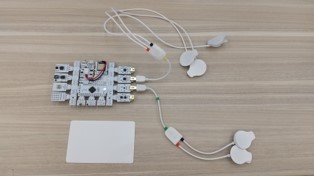}
    		\subcaption{Wearable biometric sensor.}
    	\end{minipage}\hfill
    	\begin{minipage}{0.45\linewidth}
    		\centering
    		\includegraphics[width=0.99\linewidth]{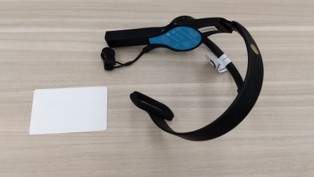}
    		\subcaption{Brain-wave headset.}
    	\end{minipage}\hfill
	\end{minipage}\vfill
	\caption{Sensors used in experiment.}
	\label{fig:sensors}
\end{figure}

We use a video camera, which does not require any additional effort from workers for sensing. 
In addition, we use wearable ECG, EEG, EOG, eye-tracking sensor, and accelerometer since they were found to be useful for estimating internal states of humans in the previous studies reviewed in section \ref{sec:related_work}.
GSR sensors were also found to be useful, but we did not use one since we found it significantly interfered with the task in a preliminary experiment.
The sensors used in the experiment are Logicool HD Pro Webcam C920 (a camera for third-person-view video), SMI Eye-Tracking Glasses (ETG) (a glasses-type camera for first-person-view video and gaze information), TicWatch Pro (a smartwatch for acceleration data and heart rate data), biosignalsplux (a wearable biometric sensor for ECG and EOG data), and MindWave Mobile 2 (a mobile brain-wave headset for EEG data).
Figure \ref{fig:sensors} shows these sensors.

\section{\uppercase{Estimation Method}}

\subsection{Feature Extraction}\label{sec:feature_extraction}
This section explains the feature extraction method for the human-error-potential estimation.
First, section~\ref{sec:overall} outlines the overall strategy for feature extraction.
Then, section~\ref{sec:noise} explains how to robustly calculate important biometric indices using noisy sensor data.
Finally, section~\ref{sec:action} introduces the features that represent the body movement.

\subsubsection{Overall strategy}\label{sec:overall}
The raw signals of ECG, EOG, and EEG data are not usually used for the analysis.
Alternatively, certain kinds of processing such as peak detection and frequency domain analysis are first applied, and higher level indices such as heart rate, blink frequency, and signal intensity of a specific frequency of brain waves are commonly used.
This section outlines how we process the data from sensors introduced in section~\ref{sec:sensor} to calculate such basic biometric indices.
The detailed formulation for dealing with sensor noise will be given in section~\ref{sec:noise}.

\begin{figure}[t]
	\begin{minipage}{0.48\linewidth}
		\centering
		\includegraphics[width=0.99\linewidth]{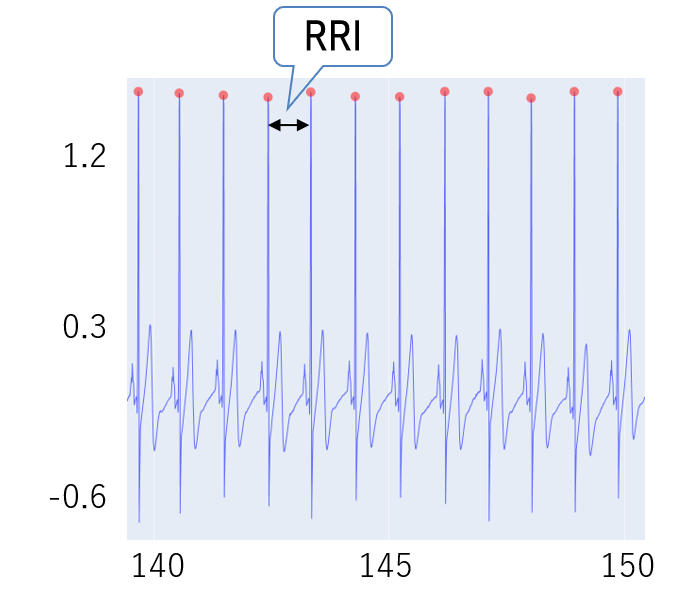}
		\subcaption{Example of ECG data.}
	\end{minipage}\hfill
	\begin{minipage}{0.47\linewidth}
		\centering
		\includegraphics[width=0.9\linewidth]{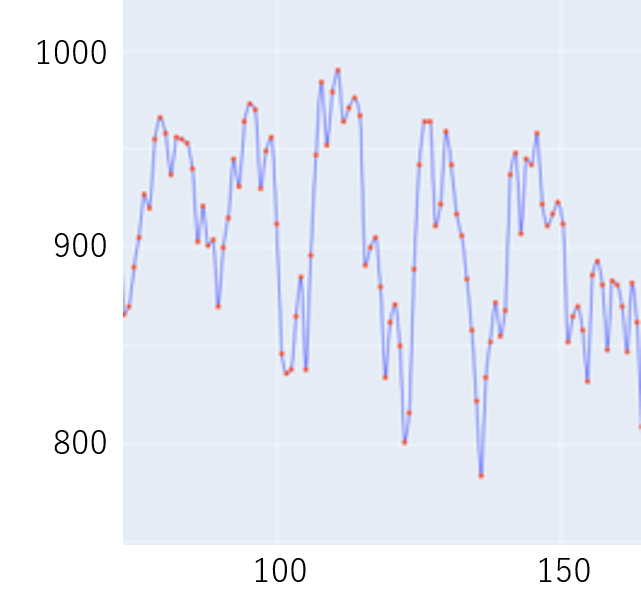}
		\subcaption{Example of RRI.}
	\end{minipage}\hfill
	\caption{Example of ECG data and its RR interval (when a subject is staying calm). x axis: time (s), y axis: electric potential (mV) for (a) and interval (ms) for (b).}
	\label{fig:calm}
\end{figure}

ECG data usually have periodic cycles that correspond to the heart beat as shown in Figure~\ref{fig:calm}(a).
We first detect peaks, called R waves represented by red circles in the figure, and calculate the intervals of the R waves, which are called RR intervals, or RRI for short (Figure~\ref{fig:calm}(b)).
We then extract the following ECG-related features per trial: 
the mean RRI over each trial, the strength of low frequency signals (4 Hz $\le~f < $ 15 Hz) of RRI wave called LF, that of high frequency signals (15 Hz $\le~f$) called HF, and the ratio LF/HF.
HF and LF are widely used as indicators of the cardiac parasympathetic nerve activity and that of both parasympathetic and sympathetic components, respectively.

We apply the similar feature extraction to smartwatch data.
TicWatch Pro provides heart rate data, which correspond to 60/RRI, every second.
Therefore, we calculate RRI on the basis of the heart rate data and extract the mean RRI, LF, HF, and LF/HF in the same way as described above.

EOG data show sharp peaks when a subject blinks.
Therefore, we first detect these peaks and use the average frequency of blinks (times/minute) over each trial as a feature.
The details of the peak detection method will be given in section \ref{sec:noise}.

The SMI ETG provides various eye-activity information including the gaze location in each video frame, and the eye-event information such as visual intake, saccade, and blink.
We extract the following features:
the standard deviation of the gaze location in horizontal and vertical directions, the mean and the standard deviation of the distances of gaze location between the two consecutive frames, the frequency of the distance being larger than a threshold, the ratio of visual-intake event, and the ratio of saccade event.
We do not use the blink event information as we found in a preliminary experiment that it was not accurate.

The API of Mindwave Mobile 2 provides the signal strength of $\delta$ wave (1-3 Hz), $\theta$ wave (4-7 Hz), low-$\alpha$ wave (8-9 Hz), high-$\alpha$ wave (10-12 Hz), low-$\beta$ wave (13-17 Hz), high-$\beta$ wave (18-30 Hz), low-$\gamma$ wave (31-40Hz), mid-$\gamma$ wave (41-50 Hz), the score of concentration, and the score of meditation.
These measurements are provided every second.
We use the mean and standard deviation over each trial as the features of EEG data.

The acceleration data obtained by the smartwatch are first converted into a movement feature using the method described in section~\ref{sec:action}.
Then we calculate its mean over each trial.
Additionally, we subtract the mean norm of the acceleration data from the raw acceleration data and then convert the result and calculate the mean in the same way as above.
This is for getting rid of the factor of gravity acceleration.

The video data of a fixed camera are first converted into movement features in a way that is described in section~\ref{sec:action} and the mean over each trial is used as video features.

Finally, we convert all the features described above into the deviation from the values in the calm state by subtracting the mean values of each feature in the calm state.
This is for reducing the between-subject bias.
The effect of this pre-processing will be discussed in section~\ref{sec:result}.

\subsubsection{Calculation of basic biometric indices under noise}\label{sec:noise}
\begin{figure}[t]
	\centering
	\includegraphics[width=0.75\linewidth]{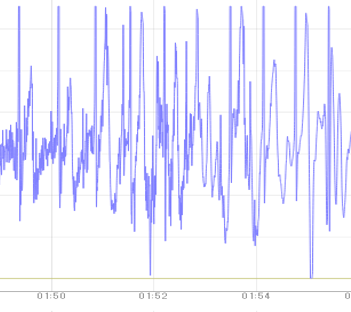}
	\caption{Example of ECG data in SSD-replacement task.}
	\label{fig:move}
\end{figure}

Wearable biometric data such as ECG and EOG data easily suffer from noise caused by body movement.
For example, ECG data show clear peaks (R wave) when a subject stays calm (Figure~\ref{fig:calm}(a)), but it becomes difficult to accurately detect an R wave when a subject moves as shown in Figure~\ref{fig:move}.
We propose a method for robustly calculating biometric indices such as RRI by formulating the calculation process using a probabilistic model, which can take the prior knowledge about each biometric index into account.
We explain the method by using calculation of RRI based on ECG data as an example.

Let $x_t$ denote the ECG measurement at time $t$ and define a set of the measurement as $X_{t_s}^{t_e} \equiv \{x_t\}_{t=t_s}^{t_e}$.
Let $t^{(n)}$ denote the timestamp of observing the $n$-th R wave, and let $y^{(n)}$ denote the $n$-th RRI value.
Note that $y^{(n)} = t^{(n+1)} - t^{(n)}$ holds by the definition of RRI.
Let $T$ be the length of ECG measurement sequence, then the $(n+1)$-th RRI given a sequence of ECG measurement $X_{1}^{T}$ and RRIs up to the $n$-th is modeled as follows.
\begin{align}
	\hspace{-50pt}p(y^{(n+1)}|X^T_1, y^{(n)}, ..., y^{(1)}) \nonumber\\
	&\hspace{-90pt}= p(y^{(n+1)}|X_{t^{(n+1)}}^{T}, y^{(n)}, ..., y^{(1)})\label{eq:post}\\
	&\hspace{-90pt}= \frac{1}{Z}p(X_{t^{(n+1)}}^{T}|y^{(n+1)}, y^{(n)}, ..., y^{(1)})\label{eq:post2}\\
	 &\hspace{-50pt}p(y^{(n+1)},y^{(n)}, ..., y^{(1)}) \nonumber\\
	&\hspace{-90pt}= \frac{1}{Z'}p(X_{t^{(n+1)}}^{T}|y^{(n+1)}, y^{(n)}, ..., y^{(1)})\label{eq:post3}\\
	 &\hspace{-50pt}p(y^{(n+1)}|y^{(n)}, ..., y^{(1)}).\nonumber
\end{align}
In equation \eqref{eq:post}, we used the assumption that the RRIs up to the $(n-1)$-th do not matter if the $n$-th R wave is given.
$Z$ and $Z'$ in equations~\eqref{eq:post2} and \eqref{eq:post3}, respectively, are constant values for normalization.

$p(X_{t^{(n+1)}}^{T}|y^{(n+1)}, y^{(n)}, ..., y^{(1)})$ in equation~\eqref{eq:post3} denotes the probability of observing a sequence of ECG measurement $X_{t^{(n+1)}}^{T}$ when an R wave, or peaks of ECG data, are observed at time $\hat{t}^{(n+1)} = t^{(1)}+\sum_{i=1}^{n+1}y^{(i)}$.
This probability is expected to be higher if the ECG measurement at time $\hat{t}^{(n+1)}$ is high.
Therefore, we model it as follows.
\begin{equation}
	p(X_{t^{(n+1)}}^{T}|y^{(n+1)}, y^{(n)}, ..., y^{(1)}) = \frac{1}{C_1}(x_{\hat{t}^{(n+1)}})^{\alpha},
\end{equation}
where $C_1$ is a normalization constant, and $\alpha$ is a hyperparameter.
$p(y^{(n+1)}|y^{(n)}, ..., y^{(1)})$ in equation~\eqref{eq:post3} represents the probability of the $(n+1)$-th RRI value being $y^{(n+1)}$ given RRIs up to the $n$-th, and we model it as follows.
\begin{align}
	&\hspace{-207pt}p(y^{(n+1)}|y^{(n)}, ..., y^{(1)})\nonumber\\
	\hspace{-0pt}= \frac{1}{C_2}\left(\mathcal{N}(y^{(n)}, \sigma_1^2)+\beta\mathcal{N}(\mu^{(n)}, \sigma_2^2)+\gamma g(y^{(n+1)})\right),\label{eq:y}
\end{align}
where $\mathcal{N}(y^{(n)}, \sigma_1^2)$ and $\mathcal{N}(\mu^{(n)}, \sigma_2^2)$ denote a normal distribution with mean $y^{(n)}$ and variance $\sigma_1^2$, and that of mean $\mu^{(n)}=\frac{1}{n}\sum_{i=1}^{n}y_i$ and variance $\sigma_2^2$, respectively.
They model the prior knowledge that RRIs usually do not drastically change compared with the previous observation, and the mean value of the past observations.
$C_2$ is a normalization constant, and $\beta$ and $\gamma$ are hyperparameters.
We define $g(y^{(n+1)})$ as follows.
\begin{equation}
	g(y^{(n+1)}) = \frac{1}{\sum_{f}h_{X_1^{t^{(n)}}}(f)}h_{X_1^{t^{(n)}}}\left({\frac{1}{y^{(n+1)}}}\right),
	\label{eq:fft}
\end{equation}
where $h_{X_1^t}(f)$ represents the signal strength of frequency $f$ in sequential data $X_1^t$, which we model by fast Fourier transform (FFT) applied to $X_1^t$.

Finally, the $(n+1)$-th RRI $y^{(n+1)}$ is estimated as follows.
\begin{equation}
	y^{(n+1)} = \argmax_{\hat{y}^{(n+1)} \in Y^{(n+1)}}{p(\hat{y}^{(n+1)}|X^T_1, y^{(n)}, ..., y^{(1)})},
\end{equation}
where $Y^{(n+1)}=\{y^{(n+1)}|y_{min} < y^{(n+1)} < y_{max} \cap x_{t^{(n)}}+y^{(n+1)} \in \mathbb{M} \}$.
Here $y_{min}, y_{max}$ are the minimum and maximum possible RRI values that we preliminarily define, 
and $\mathbb{M}$ is a set of local maxima of ECG values.

We use the same formulation for blink detection with EOG data except that we introduce a uniform distribution for equation \eqref{eq:y} since it is not reasonable to assume strong periodicity in blink detection.

\subsubsection{Extraction of movement feature}\label{sec:action}
We extract features related to subjects' body movement using acceleration data of smartwatch and video data of a fixed camera, and use them in the human-error-potential estimation method so that the method can take the body movement into account.

Let $(a_{x, t}, a_{y, t}, a_{z, t})$ denote the readings of a three-axis accelerometer of a smartwatch at time $t$.
We define the movement features calculated on the basis of acceleration data as follows.
\begin{equation}
	m^{(acc)}_t = \sqrt{a_{x, t}^2 + a_{y, t}^2 + a_{z, t}^2}
\end{equation}

To extract video-based movement features, we first use a method proposed by Pavllo~\etal~\cite{pavllo2019} to acquire a set of 3D locations of each body joint $\{(\bm{l}_t^{(1)}, ..., \bm{l}_t^{(J)})\}_{t=1}^T$, where $\bm{l}_t^{(i)}$ represents the $i$-th body joint's 3D coordinate $(l_{x,t}^{(j)}, l_{y,t}^{(j)}, l_{z,t}^{(j)})$ at time $t$.
The 3D coordinates are relative to the root joint.
We set $J$ to be 17, which is a default value in the previous work.
We define the movement features calculated on the basis of video data from a fixed camera as follows.
\begin{align}
	&\hspace{-210pt}m^{(video, j)}_t = \nonumber\\
	\sqrt{\left(l_{x,t}^{(j)} - l_{x,t-1}^{(j)}\right)^2 + \left(l_{y,t}^{(j)} - l_{y,t-1}^{(j)}\right)^2 + \left(l_{z,t}^{(j)} - l_{z,t-1}^{(j)}\right)^2}
\end{align}
Note that these features are calculated joint-wise, which results in $J$ features being acquired.

\subsection{Feature Selection}\label{sec:feature_selection}
\begin{figure}
	\begin{minipage}{\linewidth}
    	\begin{minipage}{0.5\linewidth}
    		\centering
    		\includegraphics[width=0.99\linewidth]{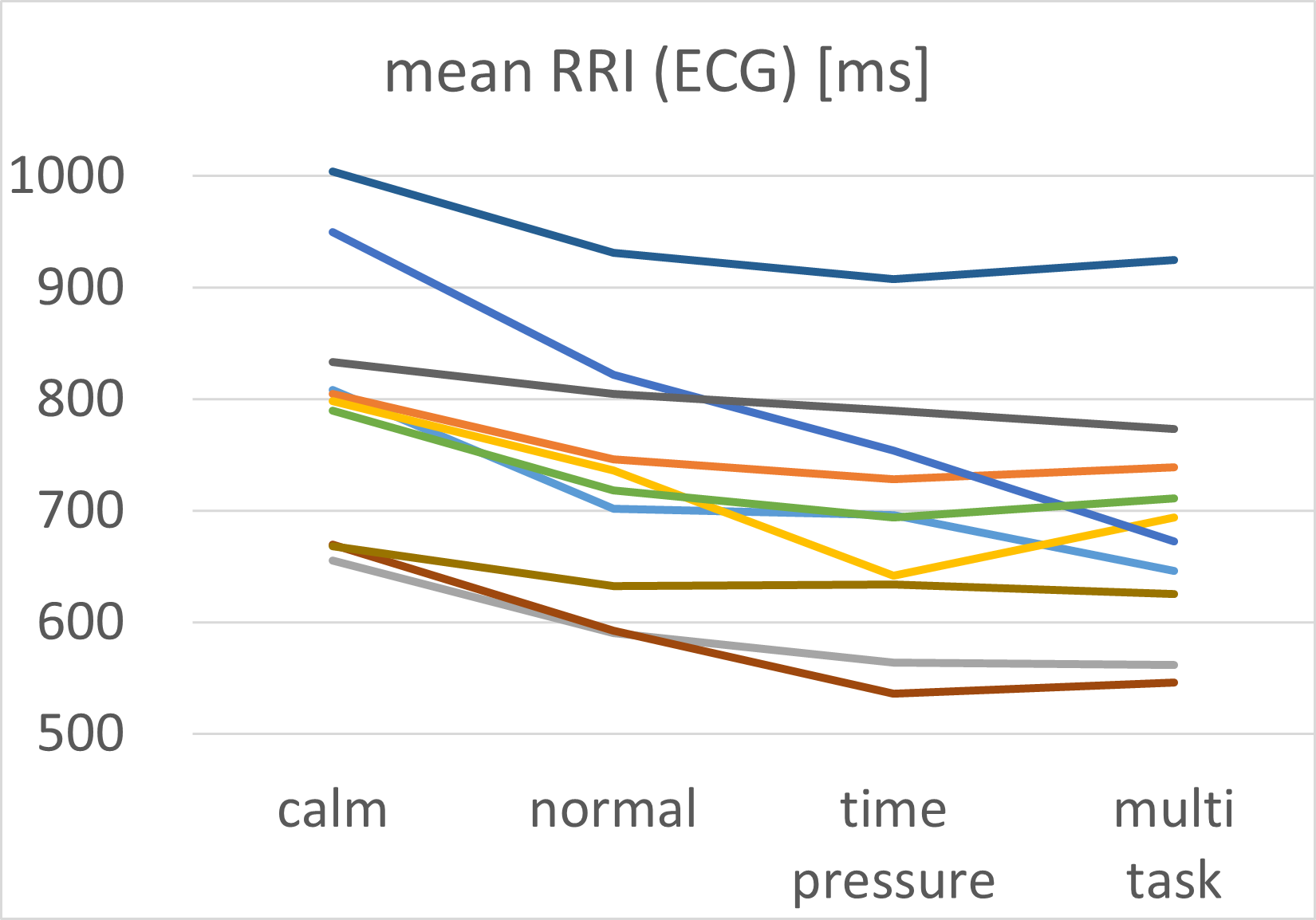}
    	\end{minipage}\hfill
    	\begin{minipage}{0.5\linewidth}
    		\centering
    		\includegraphics[width=0.99\linewidth]{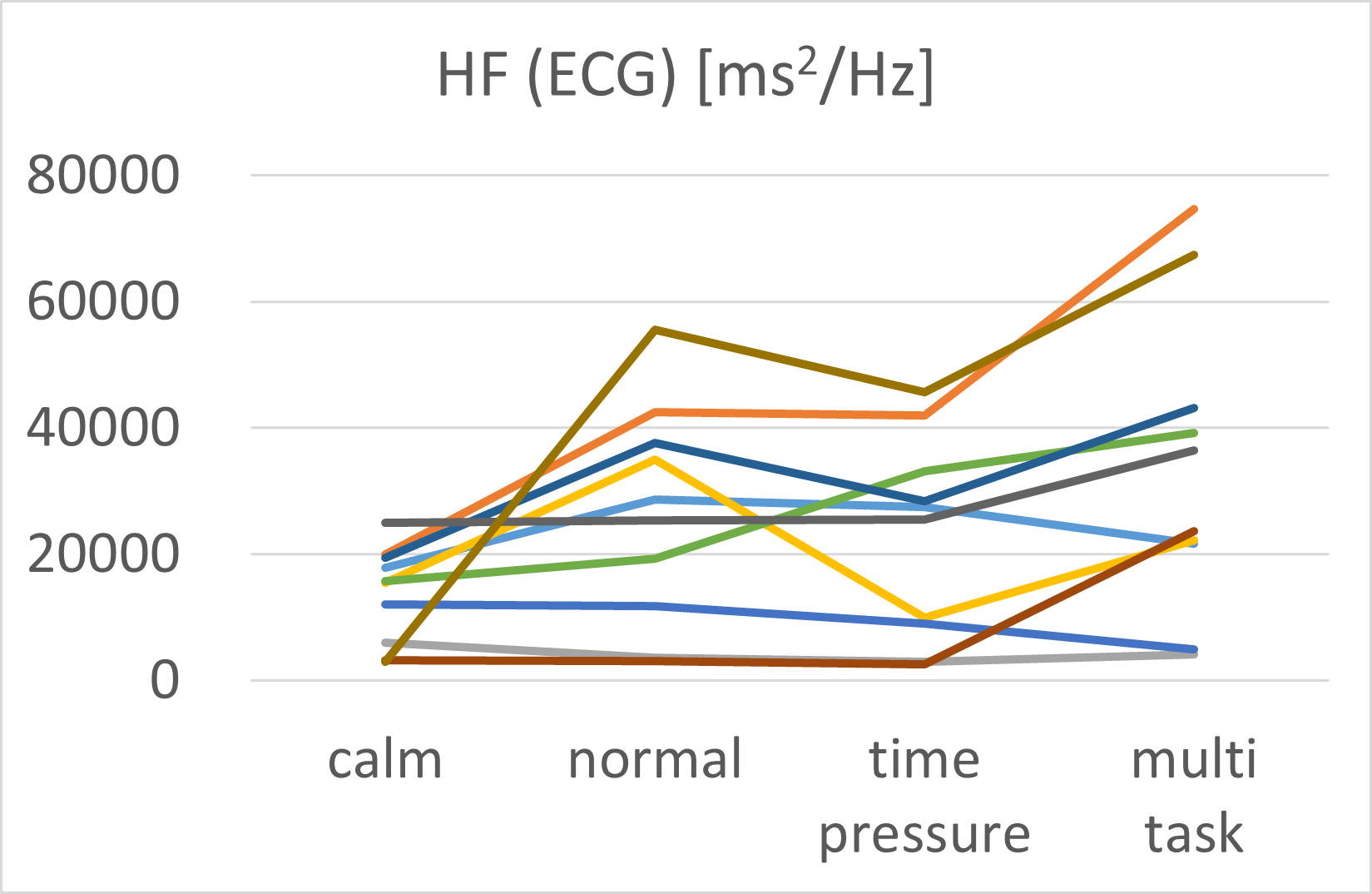}
    	\end{minipage}\hfill
	\end{minipage}\vfill
	\begin{minipage}{\linewidth}
    	\begin{minipage}{0.5\linewidth}
    		\centering
    		\includegraphics[width=0.99\linewidth]{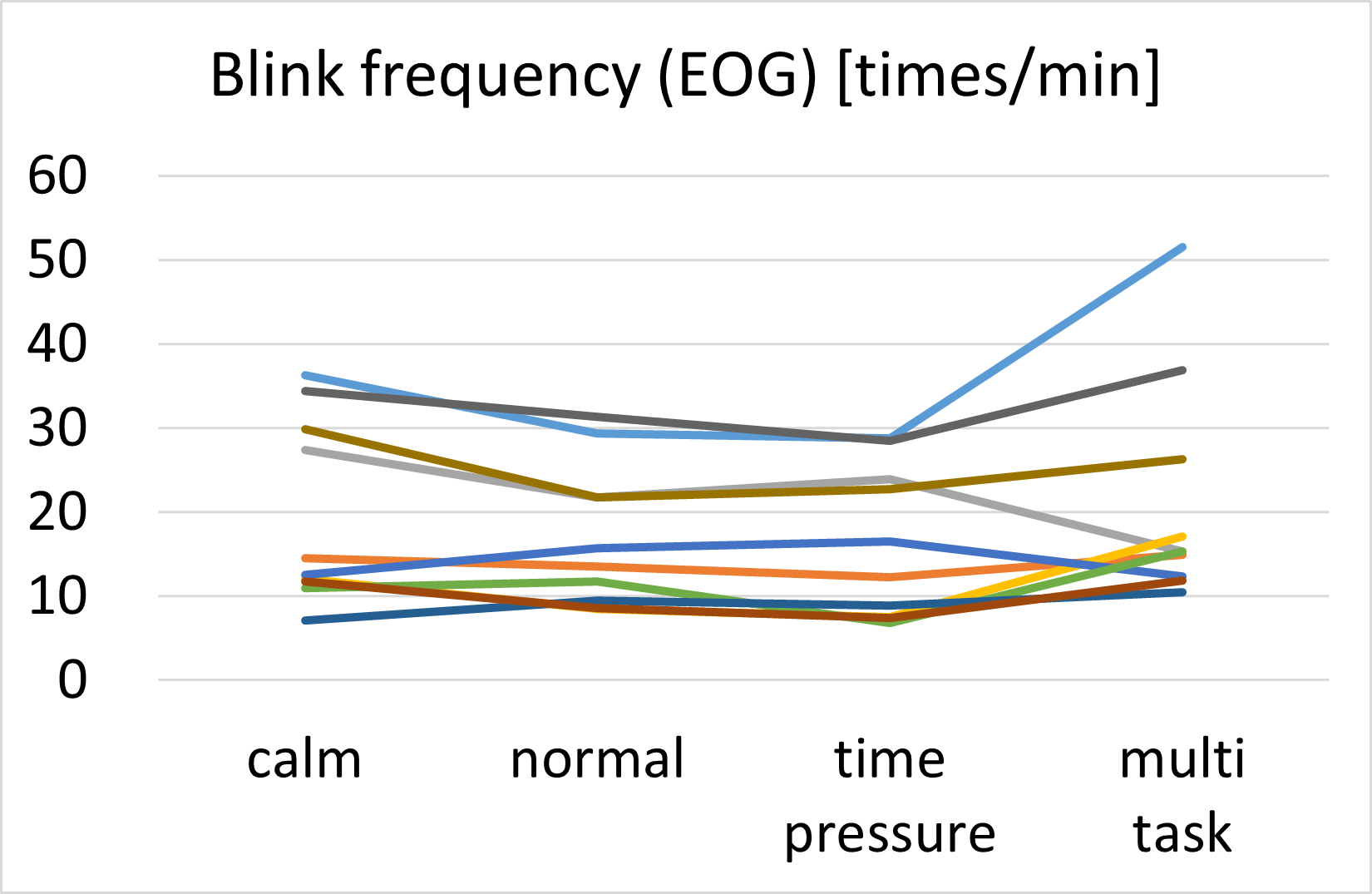}
    	\end{minipage}\hfill
    	\begin{minipage}{0.5\linewidth}
    		\centering
    		\includegraphics[width=0.99\linewidth]{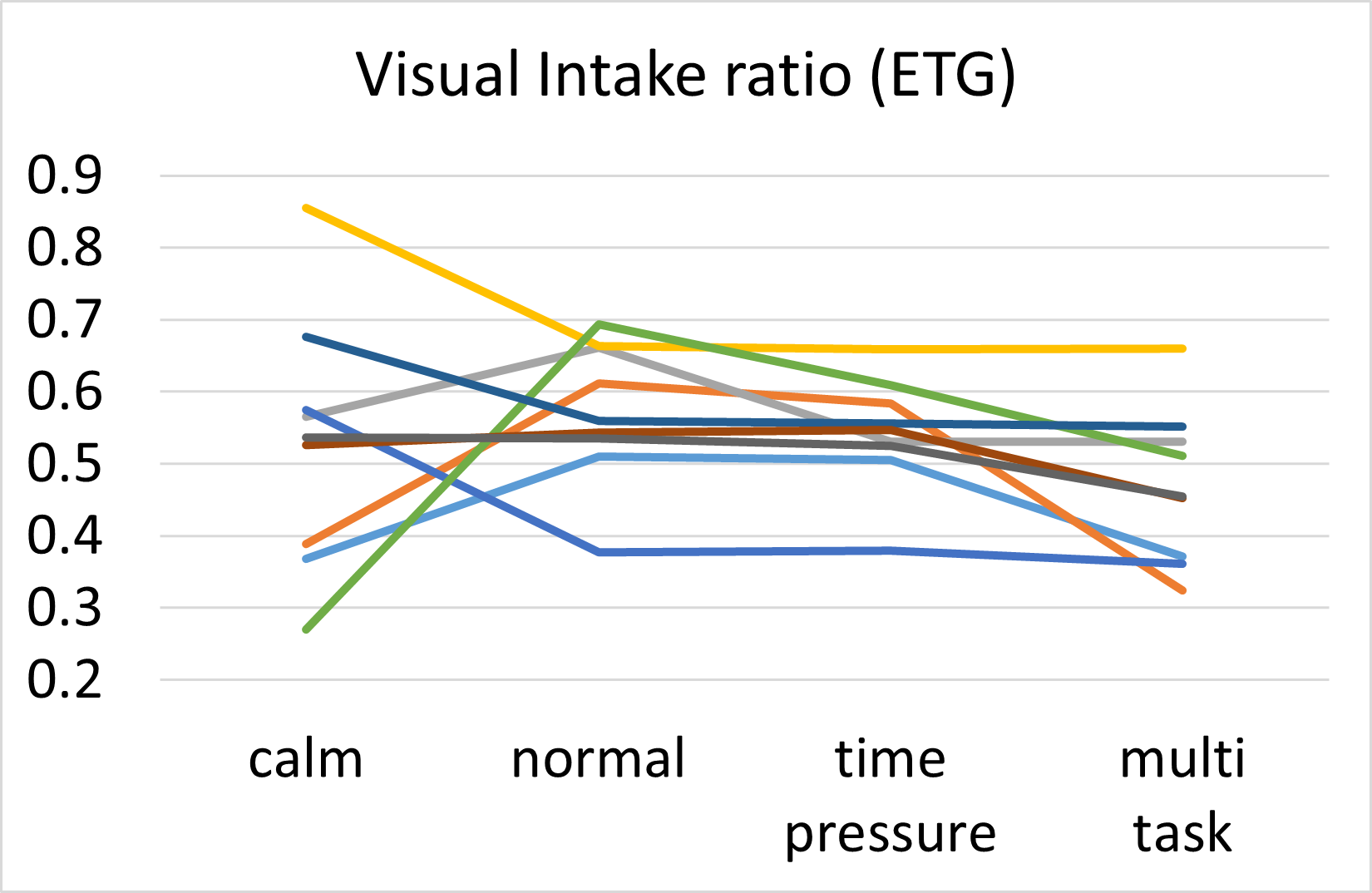}
    	\end{minipage}\hfill
	\end{minipage}
	\begin{minipage}{\linewidth}
    	\begin{minipage}{0.5\linewidth}
    		\centering
    		\includegraphics[width=0.99\linewidth]{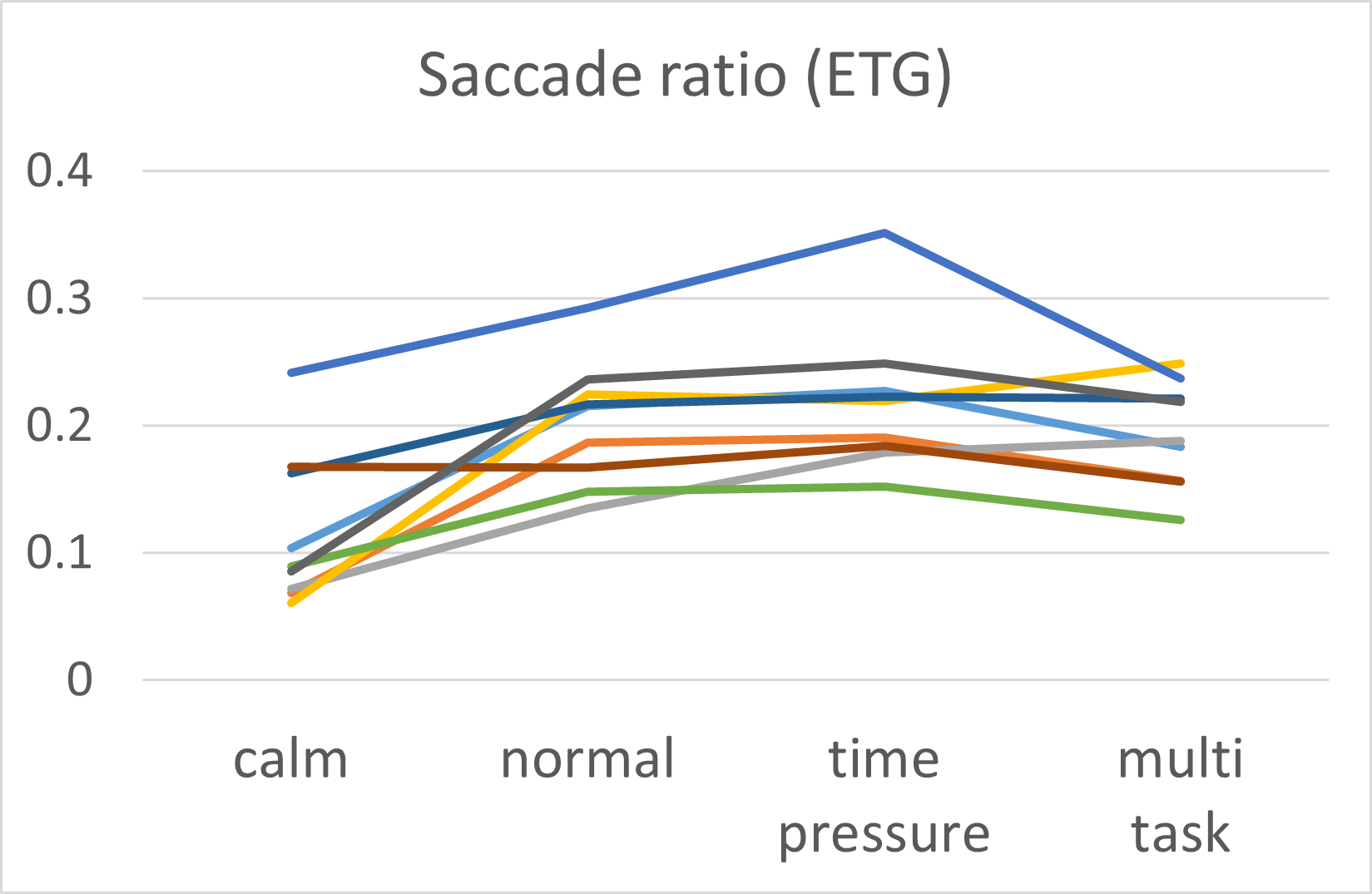}
    	\end{minipage}\hfill
    	\begin{minipage}{0.5\linewidth}
    		\centering
    		\includegraphics[width=0.99\linewidth]{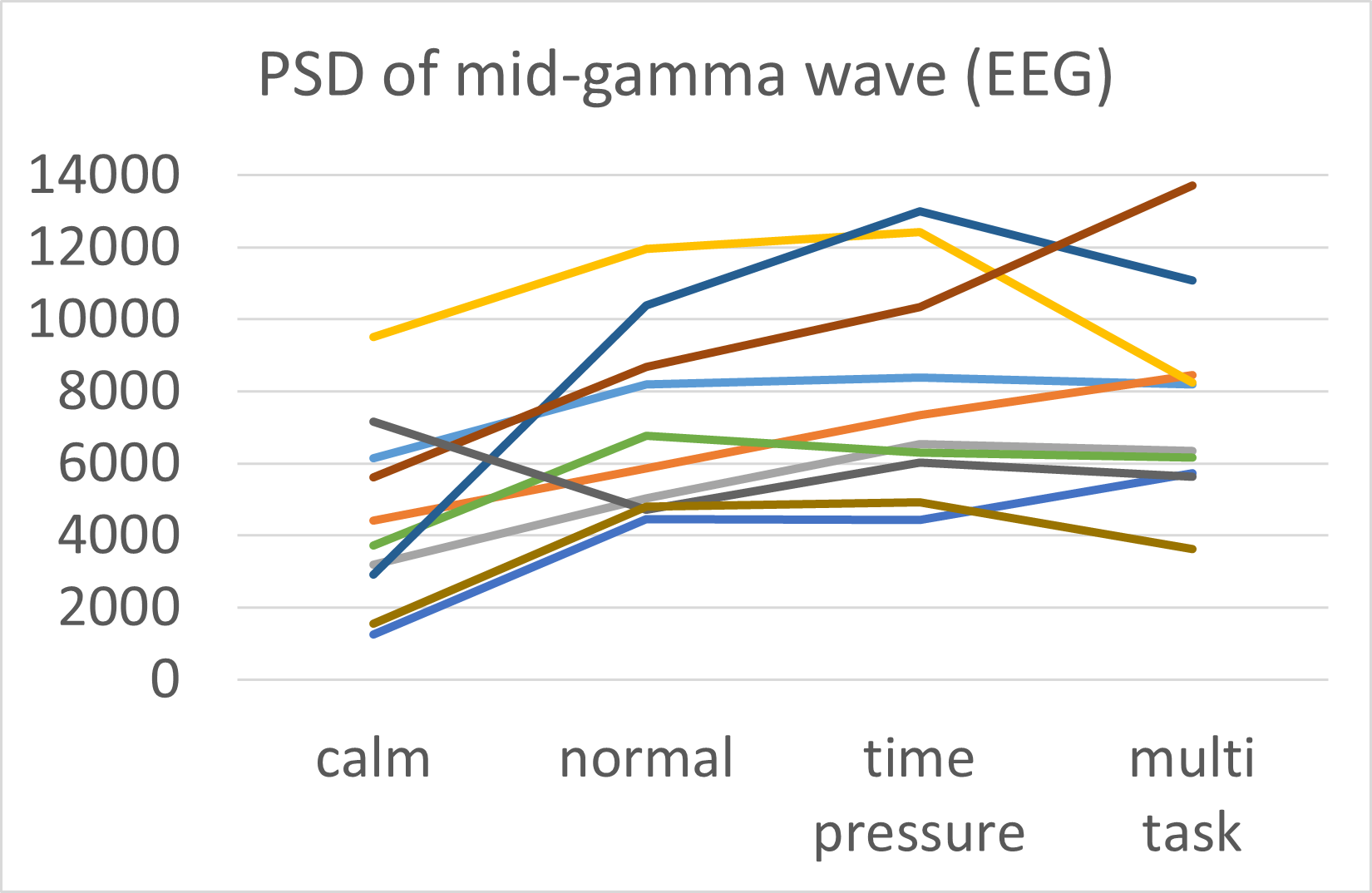}
    	\end{minipage}\hfill
	\end{minipage}
	\begin{minipage}{\linewidth}
    	\begin{minipage}{0.5\linewidth}
    		\centering
    		\includegraphics[width=0.99\linewidth]{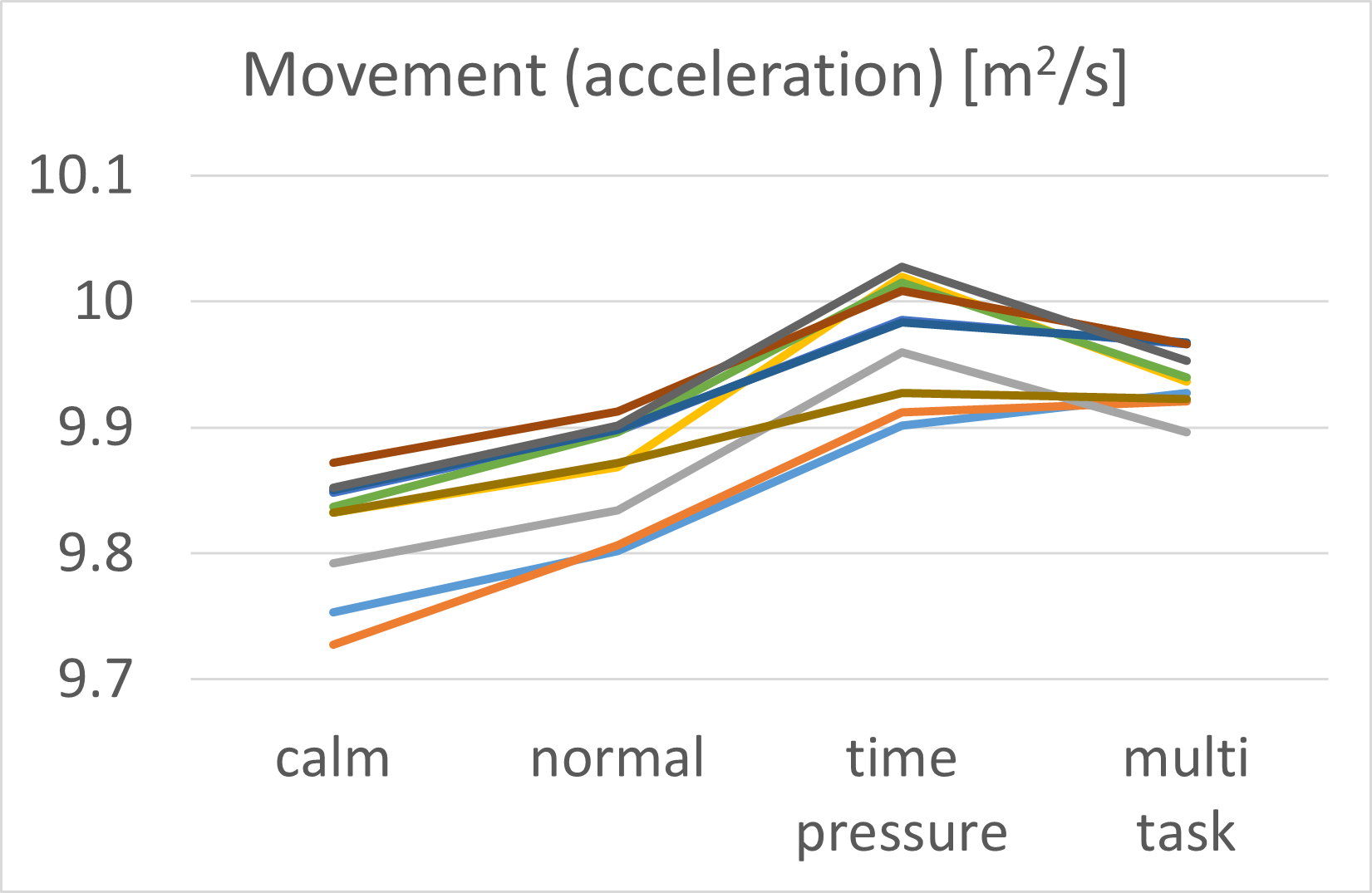}
    	\end{minipage}\hfill
    	\begin{minipage}{0.5\linewidth}
    		\centering
    		\includegraphics[width=0.99\linewidth]{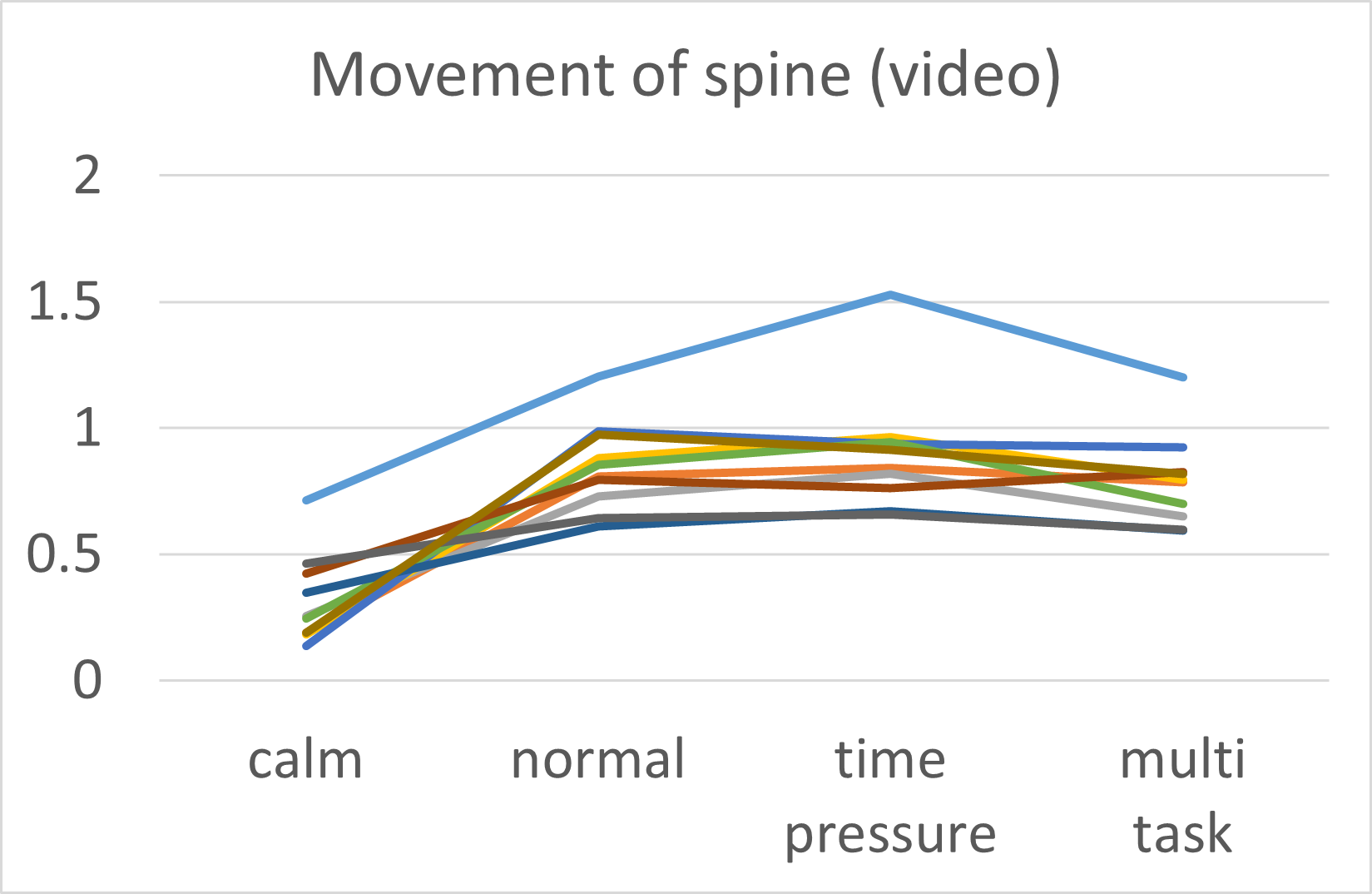}
    	\end{minipage}\hfill
	\end{minipage}
	\begin{minipage}{\linewidth}
    	\begin{minipage}{0.5\linewidth}
    		\centering
    		\includegraphics[width=0.99\linewidth]{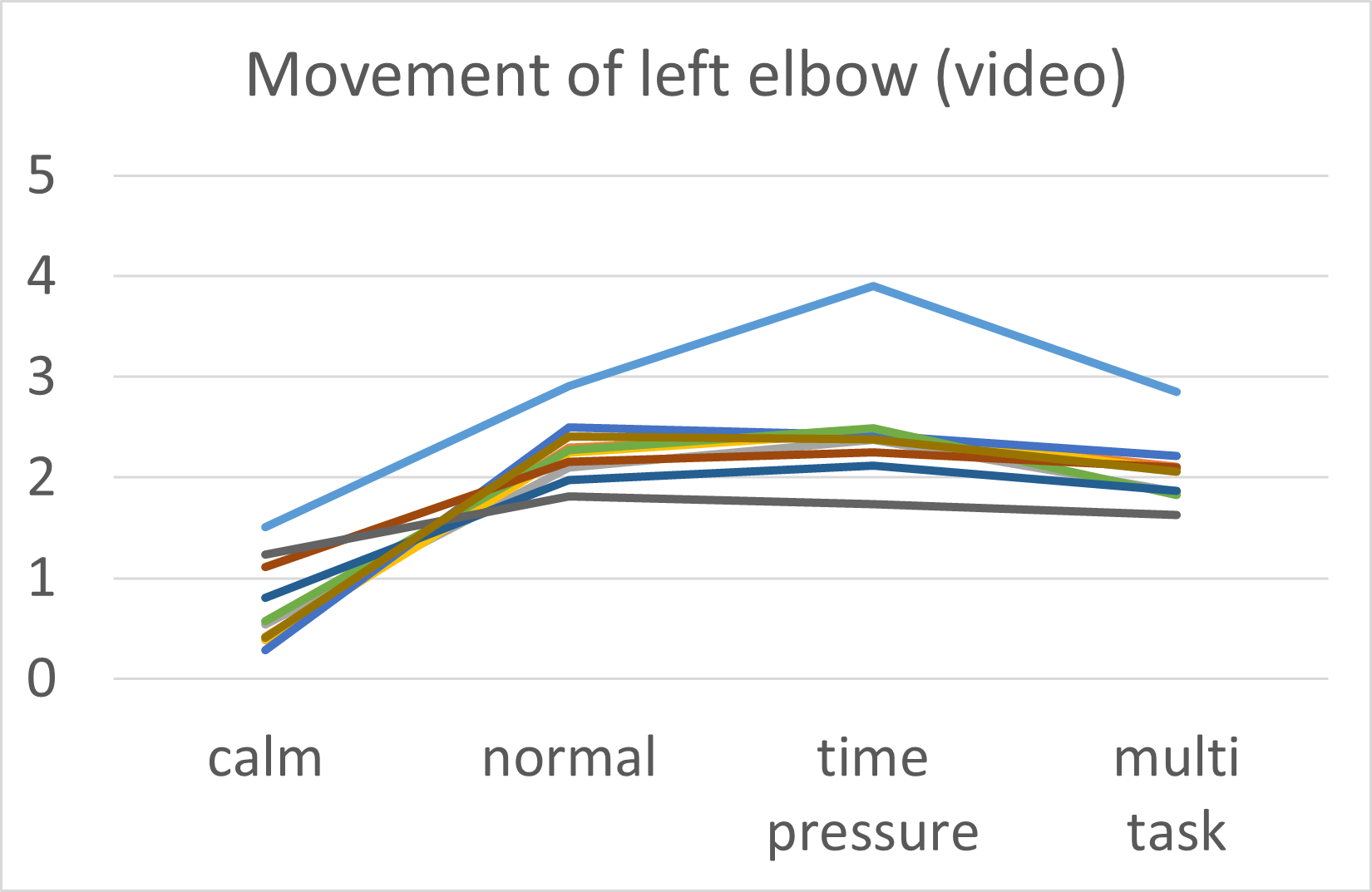}
    	\end{minipage}\hfill
    	\begin{minipage}{0.5\linewidth}
    		\centering
    		\includegraphics[width=0.99\linewidth]{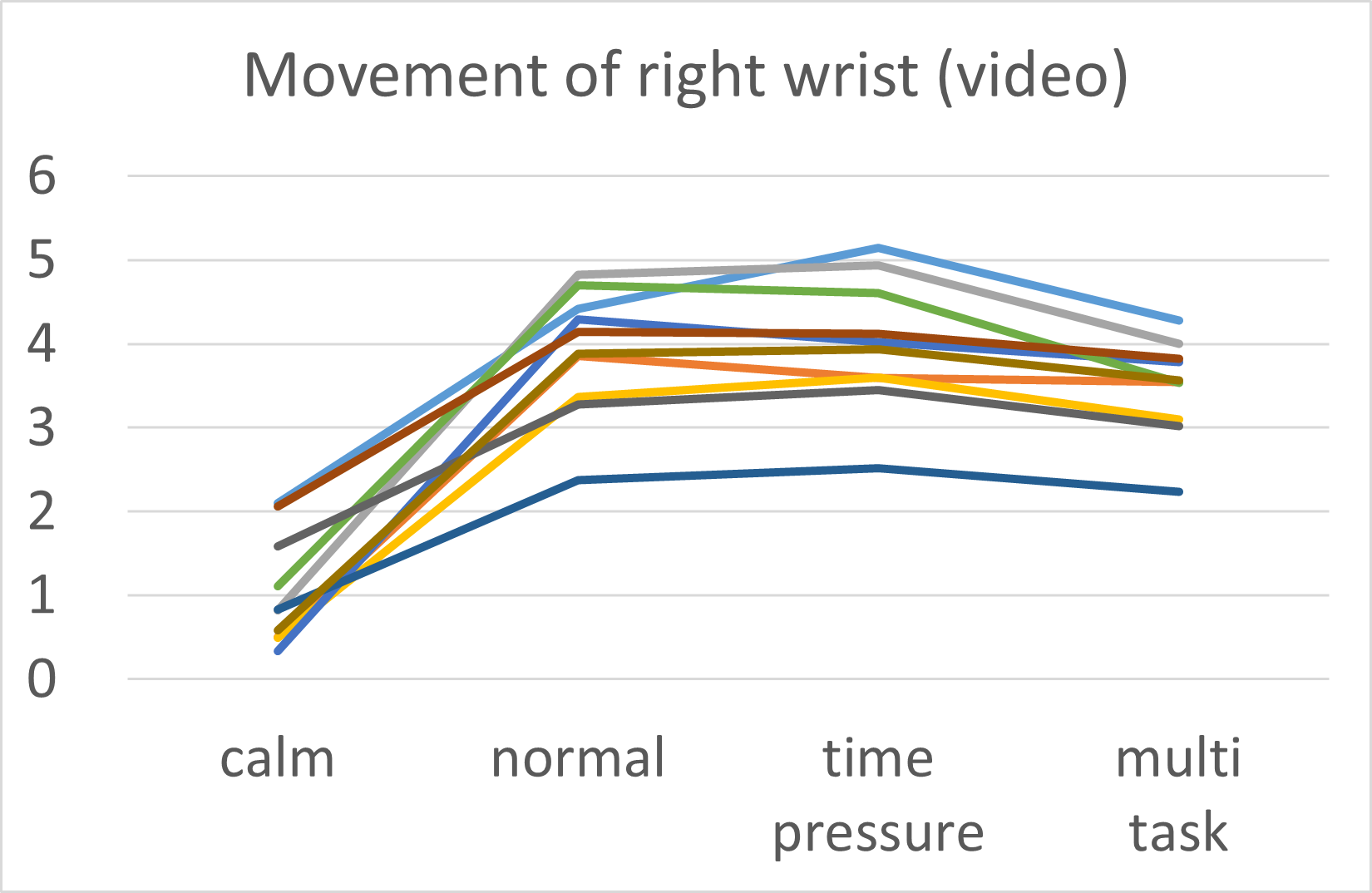}
    	\end{minipage}\hfill
	\end{minipage}
	\caption{Ten selected features.The different colors correspond to different subjects. Note that visual intake ratio and saccade ratio have only nine subjects' data because a subject could not wear ETG due to an eyesight problem.}
	\label{fig:features}
\end{figure}

We can obtain 55 different features in total by the feature extraction method described in section~\ref{sec:feature_extraction}.
Although it may be possible to make the model learn sufficiently well using all the 55 features if there are enough training data, the model may suffer from over-fitting if the available training data are limited.
Therefore, we analyzed each feature in detail and selected a set of features that showed a noticeable tendency depending on the task conditions (normal, time-pressure, and multi-task).
As a consequence of the analysis, we ended up selecting the 10 features shown in Figure~\ref{fig:features}.

\subsection{Classification Model}\label{sec:classification}
We use neural network for the classification model unless otherwise stated.
We use a simple fully connected network rather than convolutional neural networks (CNN) and recurrent neural networks (RNN) since it is not reasonable to assume locality and other specific dependencies among the features described in section~\ref{sec:feature_selection}.
We will discuss the models other than neural networks in section~\ref{sec:result}.

\section{\uppercase{Evaluation}}
\subsection{Experimental Settings}\label{sec:exp_setting}
We conducted an experiment to verify the effectiveness of the proposed method with 10 subjects.
The experiment was conducted in a lab environment with lab members.
The experiment on a real shop floor with real workers is one of our future works.

\begin{figure*}[t]
	\centering
	\includegraphics[width=0.99\linewidth]{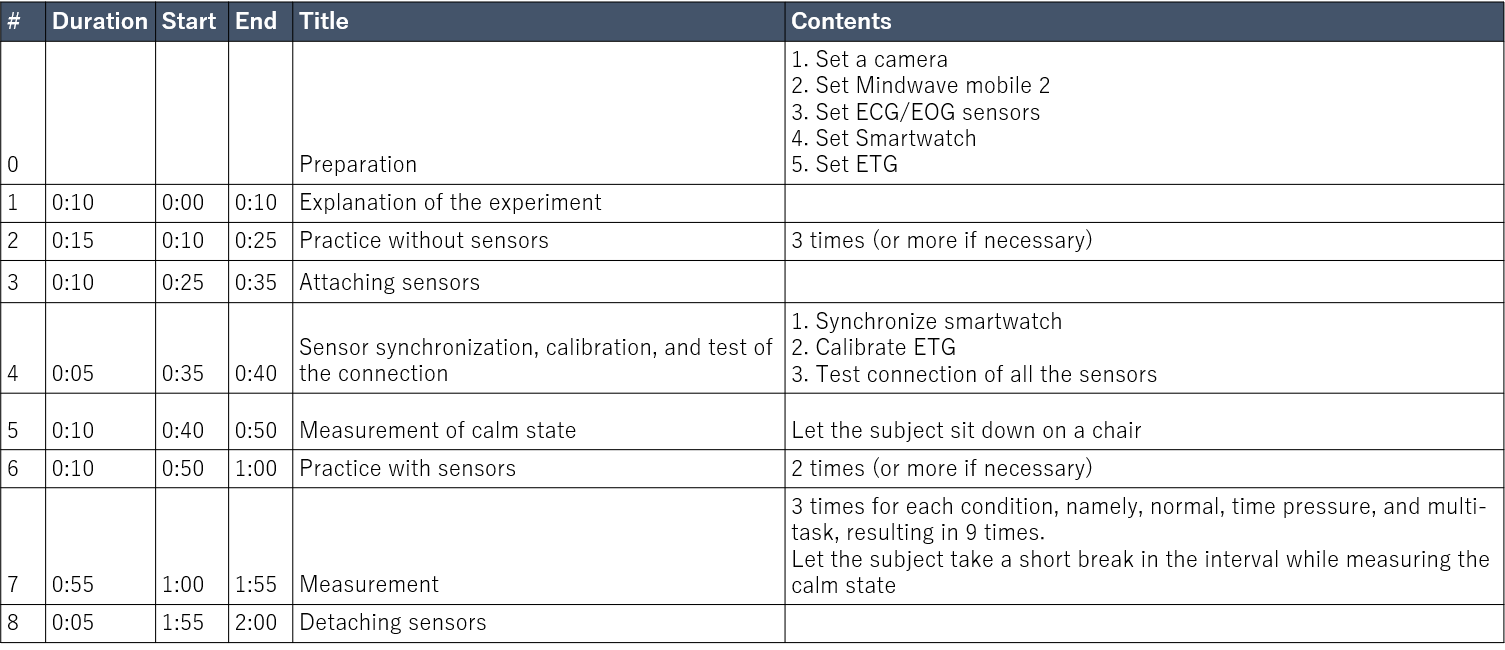}
	\caption{Experimental procedure.}
	\label{fig:protocol}
\end{figure*}

The overall experimental procedure is shown in Figure~\ref{fig:protocol}.
First, we explained the experiment to the subjects.
Then to eliminate the learning effect, they practiced the SSD-replacement task without attaching any sensors until they were sufficiently familiar with the task.
After they become confident doing the task, we attached the wearable sensors and synchronized all the sensors, calibrated the sensors if necessary, and tested the connection.
Then we let the subjects be seated on chairs and asked them to stay calm.
We collected sensor data during this calm state.
Then the subjects practiced the task again with sensors attached and confirmed that the sensors did not disturb them in the task.
We collected three trials per condition (normal, time-pressure, multi-task), resulting in nine trials per subject in total.
We let the subjects take a short break after every trial.
We collected sensor data in this break time as well for using them as measurement in the calm state.
After finishing the first trial (\eg normal condition) and the subsequent short break, subjects worked on another condition (\eg time-pressure) and took a short break again.
Then they worked on the remaining condition (\eg multi-task) and took a short break.
Hereinafter, we call this set of three sequential trials a ``{\it cycle}''.
They repeated this cycle three times.
We let a one-third of the subjects start with the normal condition, another third start with the time-pressure condition, and the rest start with the multi-task condition for the counterbalance.
One trial usually took from two to four minutes, and the whole experiment including the preparation took approximately two hours per subject.

\subsection{Evaluation Protocol}\label{sec:evaluation_protocol}
We evaluated the performance by three-fold cross validation, in which we split the data by aforementioned cycles, unless otherwise stated.
More specifically, we used the data from two cycles for training and used the data from the other cycle for test, and repeated this three times with different combinations of training and test data.
This resulted in 6 trials from each subject for training data and the remaining 3 trials from each subject for test data, ending up with 60 trials in total for training data and 30 trials in total for test data.
We used the mean accuracy of this three-fold cross validation for the basic evaluation metric.

\subsection{Implementation details}\label{sec:imple}
We optimized the hyperparameters including learning rate, learning schedule, number of layers, and number of nodes using the TPE algorithm~\cite{bergstra2011} with Optuna~\cite{akiba2019optuna}.
As a result, we set the number of hidden layers to 3 and the number of nodes in each layer to 200, 50, and 50.
The initial learning rate was 0.03, and it was decayed by multiplying $1/t^{0.5}$ at the $t$-th epoch.
The hyperparameters of models other than neural networks compared in Table~\ref{tab:different_models} were also optimized in the same way.
If a feature value was missing, we interpolated it with the median value of the feature value over training data.
All the features were normalized by subtracting means of each feature in the training data and dividing them by standard deviations.

\subsection{Results and Discussion}\label{sec:result}
\paragraph{Main result.}
Table~\ref{tab:result} shows the main evaluation results.
The overall accuracy was 74.4\% (=67/90), and the accuracy of binary classification between normal condition and abnormal conditions, where human-error tends to occur more frequently was 82.2\% (=74/90).
The table shows that the multi-task condition was more clearly distinguishable from the normal condition than the time-pressure condition was.
This observation agrees with the fact that all the subjects said that they felt the multi-task condition was the most difficult and the normal condition was the easiest.
\begin{table}[t]
    \centering
    \caption{Evaluation result. Time: time-pressure condition, Multi: multi-task condition, GT: ground truth.}
    \label{tab:result}
    \begin{tabular}{cl|r|r|r|r|}
        \cline{3-6}
        \multicolumn{1}{l}{}                      &    & \multicolumn{4}{c|}{Estimated}                                                                               \\ \cline{3-6} 
        \multicolumn{1}{l}{}                      &    & \multicolumn{1}{l|}{Normal} & \multicolumn{1}{l|}{Time} & \multicolumn{1}{l|}{Multi} & \multicolumn{1}{l|}{Total} \\ \hline
        \multicolumn{1}{|c|}{\multirow{4}{*}{GT}} & Normal & 7.3                       & 2.3                    & 0.3                    & 10                      \\ \cline{2-6} 
        \multicolumn{1}{|c|}{}                    & Time & 2.3                    & 6.7                       & 1.0                    & 10                      \\ \cline{2-6} 
        \multicolumn{1}{|c|}{}                    & Multi & 0.3                       & 1.3                    & 8.3                    & 10                      \\ \cline{2-6} 
        \multicolumn{1}{|c|}{}                    & Total & 10.0                & 10.3                & 9.7                   & 30   \\ \hline
    \end{tabular}
\end{table}

\paragraph{Analysis on feature pre-processing.}\label{sec:preprocess}
Table~\ref{tab:preprocess} shows the effect of feature pre-processing described at the end of section~\ref{sec:overall}.
``Absolute'' denotes the result obtained by using all the features directly, whereas ``Relative'' denotes the result obtained by converting features into the deviation from the values in the calm state by subtracting the mean values of each feature in the calm state.
As the table indicates, ``Relative'' gave slightly better performance than ``Absolute''.
We think this is because different subjects have different base values for biometric indices, and using values relative to the calm states reduces this between-subject bias, ending up in enabling us to focus more on the difference in the conditions.
\begin{table}[t]
	\caption{Comparison of feature pre-processing methods.}
	\label{tab:preprocess}
	\centering
	\begin{tabular}{ccc}
		\toprule
		Pre-processing & 3 classes & 2 classes\\\midrule
		Absolute & 73.3 & 81.1\\
		Relative & 74.4 & 82.2\\
		\bottomrule
	\end{tabular}
\end{table}

\paragraph{Ablation study for the feature selection and the movement feature.}
Table~\ref{tab:ablation} shows the ablation study for the feature selection method described in section~\ref{sec:feature_selection} and the movement features described in section~\ref{sec:action}.
The effectiveness of the feature selection is demonstrated by comparing \#1 with \#2, and \#3 with \#4.
This suggests the importance of excluding unnecessary features probably because the available training data were small.
Similarly, the effectiveness of the movement feature is demonstrated by comparing \#1 with \#3, and \#2 with \#4.
This suggests the importance of taking into account the information of body movement when attempting to estimate human-error potential, and possibly other psychophysiological indices as well, in a situation where target subjects do not stay calm.
\begin{table}[t]
	\caption{Ablation for the feature selection and the movement feature. FS: the feature selection method described in section~\ref{sec:feature_selection}. If this is unchecked, results are based on all the available features without applying the feature selection. MF: the movement features described in section~\ref{sec:action}. If this is unchecked, results are only based on the biometric-sensor data without using the movement features. \#F: the number of used features. 3 classes: normal, time-pressure, and multi-task condition, 2 classes: normal condition and abnormal condition where human-error tends to occur more frequently.}
	\label{tab:ablation}
	\centering
	\begin{tabular}{cccccc}
		\toprule
		\# & FS & MF & \#F & 3 classes & 2 classes\\\midrule
		1 &            &            & 36 & 56.5 & 71.1\\
		2 & \checkmark &            & 6  & 70.0 & 78.9\\
		3 &            & \checkmark & 55 & 61.1 & 65.6\\
		4 & \checkmark & \checkmark & 10 & 74.4 & 82.2\\
		\bottomrule
	\end{tabular}
\end{table}

\paragraph{Analysis on feature-selection methods.}
Table \ref{tab:feature_selection} compares different feature selection methods.
\#1 denotes the result obtained by using all the features without applying any feature selection method.
Note that the movement features are also included.
\#2 denotes the result obtained by reducing the number of dimensions to 10 with principal component analysis (PCA).
Interestingly, \#2 did not result in better accuracy than \#1.
This may be because the neural network could learn the (linear) transform that was equivalent to PCA in the first layer.
\#3 is the result obtained by selecting features greedily (see Algorithm~\ref{alg:greedy} for details).
\#4 is the result obtained by using the manual feature selection described in section \ref{sec:feature_selection}.
Although the greedy feature selection gave slightly better results than \#1 and \#2, the analysis-based selection resulted in much better accuracy.
The result suggests the effectiveness of traditional feature engineering in this field especially when the number of candidate features is relatively low.
\begin{table}[t]
	\caption{Comparison of different feature-selection methods.}
	\label{tab:feature_selection}
	\centering
	\begin{tabular}{clccc}
		\toprule
		\# & Feature selection & \#F & 3 classes & 2 classes\\\midrule
		1 & All features & 55 & 61.1 & 65.6\\
		2 & PCA & 10 & 61.1 & 65.6\\
		3 & Greedy & 7 & 65.6 & 68.9 \\
		4 & Analysis-based & 10 & 74.4 & 82.2\\
		\bottomrule
	\end{tabular}
\end{table}
\begin{algorithm}
	\SetAlgoLined
	\DontPrintSemicolon
	\caption{Greedy feature selection.}         
	\label{alg:greedy}
	\SetKwInOut{Input}{Input}\SetKwInOut{Output}{Output}
	\SetKw{In}{in}
	\Input{\xvbox{2mm}{$F$} -- a set of features\\
		\xvbox{2mm}{$m$} -- a classification model\\
		\xvbox{2mm}{$N$} -- the maximum number of \\\hspace{20pt}features to be selected\\
		}
	\Output{$BF$ -- best features}
	\BlankLine
	$SF\leftarrow \{\}$\tcp*[f]{a set of selected features}\;
	$n\leftarrow 1$\;
	$a_{best}\leftarrow -1$\;
	\For{$n~\le~N$}{
    	$a^{(n)}_{best}\leftarrow -1$\;
	    \For{$f~\In~F$}{
	        $a\leftarrow m\left(SF\cup\{f\}\right)$\;
	        \If{$a > a^{(n)}_{best}$}{
                $a^{(n)}_{best}\leftarrow a$\;
                $f_{selected}\leftarrow f$\;
	        }
	    }
	    $SF\leftarrow SF~\cup~\{f_{selected}\}$\;
	    $F\leftarrow F\setminus~\{f_{selected}\}$\;
	    \If{$a^{(n)}_{best} > a_{best}$}{
    	    $BF\leftarrow~SF$\;
    	    $a_{best}\leftarrow~a^{(n)}_{best}$\;
	    }
	}
\end{algorithm}

\paragraph{Analysis on classification models.}\label{sec:models}
Table~\ref{tab:different_models} compares different classification models.
We found the neural network worked the best.
Note that the 10 manually selected features are used for all the models.
\begin{table}[ht]
	\caption{Comparison of different classification methods.}
	\label{tab:different_models}
	\centering
	\begin{tabular}{lcc}
		\toprule
		Classification method & 3 classes & 2 classes\\\midrule
		Gaussian Na\"ive Bayes & 55.6 & 67.8\\
		Decision Tree & 65.6 & 72.2\\
		$k$ nearest neighbor & 66.7 & 74.5\\
		Random Forest & 68.9 & 75.6\\
		Support Vector Machine & 72.2 & 81.1\\
		Neural network & 74.4 & 82.2\\
		\bottomrule
	\end{tabular}
\end{table}

\paragraph{Analysis on sensors.}
Table \ref{tab:sensors} shows the performance depending on the available sensors.
For the deployment on shop floors, subjects, or workers, should attach as few sensors as possible so that the sensors do not disturb their work.
The ideal scenario is to use only a fixed camera, which does not disturb workers at all, but this was found to be unrealistic from the viewpoint of accuracy (\#1).
The next reasonable scenario is to use a smartwatch alone (\#2) or in combination with a fixed camera (\#3).
However, the accuracies in these conditions were still not satisfactory.
Adding EOG (\#4), EEG (\#5), and ETG (\#6) one by one did not improve the performance compared with using only a fixed camera and a smartwatch.
In contrast, using an ECG sensor in combination with a fixed camera and a smartwatch achieved significantly better results.
This result suggests that it is very important to include features calculated on the basis of ECG signals such as mean RRI and HF and indirectly implies that the proposed method for calculating biometric indices under movement noise worked well.

\begin{table*}[ht]
	\caption{Comparison of performances by different combinations of sensors.}
	\label{tab:sensors}
	\centering
	\begin{tabular}{ccccccccc}
		\toprule
		\# & Fixed camera & Smart watch & EOG & EEG & ETG & ECG & 3 classes & 2 classes\\\midrule
		1 & \checkmark & & & & & & 51.1 & 66.7\\
		2 &  & \checkmark & & & & & 53.3 & 67.8\\
		3 & \checkmark & \checkmark & & & & & 56.7 & 66.7\\
		4 & \checkmark & \checkmark & \checkmark & & & & 52.2 & 67.8\\
		5 & \checkmark & \checkmark & & \checkmark & & & 53.3 & 70.0\\
		6 & \checkmark & \checkmark & & & \checkmark & & 53.3 & 72.2\\
		7 & \checkmark & \checkmark & & & & \checkmark & 68.5 & 75.8\\
		8 & \checkmark & \checkmark & \checkmark & \checkmark & \checkmark & \checkmark & 74.4 & 82.2\\
		\bottomrule
	\end{tabular}
\end{table*}

\paragraph{Analysis on the generalization to a new worker.}
To verify the generalizability of the proposed model to new workers, we evaluated the performance in leave-one-subject-out cross validation.
The model was trained with nine subjects' data and tested with the remaining subject's data.
We repeated this evaluation nine times by changing the subject whose data were used for the test.
Note that we excluded one subject's data from testing since the ETG data of the subject were completely missing.
This was because the subject could not wear the ETG due to an eye sight problem.

We report the averaged results of all the nine valid subjects in Table~\ref{tab:gen_new_worker}.
Note that the total number of trials does not add up to nine because we excluded some trials due to invalid data that could not be interpolated.
The overall accuracy decreased to 58.6\%.
This means that the features used in this study have certain dependencies on individual subjects, and it is difficult to use a model trained with one subject's data to estimate the human-error potential of another subject.
Finding more subject-independent features is one of our future works.
If this result is seen from the other side, however, it means that we may be able to achieve even better performance than the results in Table~\ref{tab:result} if we can collect sufficient training data from a worker and build a customized model for that worker.

The accuracy of the binary classification also decreased, but it stayed relatively high, which was 71.4\%, suggesting the model is somewhat effective at least for the binary classification to some extent.

\begin{table}[t]
    \centering
    \caption{Evaluation result of leave-one-subject-out cross validation. Time: time-pressure condition, Multi: multi-task condition, GT: ground truth.}
    \label{tab:gen_new_worker}
    \begin{tabular}{cl|r|r|r|r|}
        \cline{3-6}
        \multicolumn{1}{l}{}                      &    & \multicolumn{4}{c|}{Estimated}                                                                               \\ \cline{3-6} 
        \multicolumn{1}{l}{}                      &    & \multicolumn{1}{l|}{Normal} & \multicolumn{1}{l|}{Time} & \multicolumn{1}{l|}{Multi} & \multicolumn{1}{l|}{Total} \\ \hline
        \multicolumn{1}{|c|}{\multirow{4}{*}{GT}} & Normal & 1.67                       & 0.44                    & 0.56                    & 2.67                      \\ \cline{2-6} 
        \multicolumn{1}{|c|}{}                    & Time & 0.67                    & 1.22                       & 0.44                    & 2.33                     \\ \cline{2-6} 
        \multicolumn{1}{|c|}{}                    & Multi & 0.56                       & 0.56                    & 1.67                    & 2.78                      \\ \cline{2-6} 
        \multicolumn{1}{|c|}{}                    & Total & 2.89                & 2.22                & 2.67                   & 7.78   \\ \hline
    \end{tabular}
\end{table}

\section{\uppercase{Conclusions}}
\label{sec:conclusion}
This study tackled on a new problem of estimating human-error potential on the basis of wearable sensors, aiming at reducing the human errors on a shop floor.
Unlike existing studies, we have attempted to estimate the human-error potential in a situation where a target person does not stay calm, which is much more difficult as sensor noise significantly increases.
We proposed a novel formulation, in which the human-error-potential estimation problem is reduced to a classification problem, and introduced a new method that can be used for solving the classification problem even with noisy sensing data.
The experimental analysis demonstrated the effectiveness of our method for estimating the human-error potential.
In addition, we found that ECG data played an important role in estimating human-error potential.
Our future work includes generalizing out method to new workers by finding out subject-independent features.


\bibliographystyle{apalike}
{\small
\bibliography{KDIR}}

\begin{thebibliography}{}

\bibitem[Akiba et~al., 2019]{akiba2019optuna}
Akiba, T., Sano, S., Yanase, T., Ohta, T., and Koyama, M. (2019).
\newblock {Optuna: A next-generation hyperparameter optimization framework}.
\newblock In {\em International conference on knowledge discovery \& data
  mining}.

\bibitem[Bergstra et~al., 2011]{bergstra2011}
Bergstra, J., Bardenet, R., Bengio, Y., and K{\'e}gl, B. (2011).
\newblock Algorithms for hyper-parameter optimization.
\newblock In {\em 25th annual conference on neural information processing
  systems (NIPS)}.

\bibitem[Edmondson, 2004]{edmondson2004learning}
Edmondson, A.~C. (2004).
\newblock Learning from mistakes is easier said than done: Group and
  organizational influences on the detection and correction of human error.
\newblock {\em The Journal of Applied Behavioral Science}, 40(1):66--90.

\bibitem[Hale et~al., 1990]{hale1990human}
Hale, A., Stoop, J., and Hommels, J. (1990).
\newblock Human error models as predictors of accident scenarios for designers
  in road transport systems.
\newblock {\em Ergonomics}, 33(10-11):1377--1387.

\bibitem[Hawkins, 1987]{Hawkins1987}
Hawkins, F.~H. (1987).
\newblock {\em {Human Factors in Flight}}.
\newblock Ashgate Publishing.

\bibitem[Panicker and Gayathri, 2019]{panicker2019survey}
Panicker, S.~S. and Gayathri, P. (2019).
\newblock A survey of machine learning techniques in physiology based mental
  stress detection systems.
\newblock {\em Biocybernetics and Biomedical Engineering}, 39(2):444--469.

\bibitem[Pavllo et~al., 2019]{pavllo2019}
Pavllo, D., Feichtenhofer, C., Grangier, D., and Auli, M. (2019).
\newblock {3D human pose estimation in video with temporal convolutions and
  semi-supervised training}.
\newblock In {\em Conference on Computer Vision and Pattern Recognition
  (CVPR)}.

\bibitem[Ralph et~al., 2014]{ralph2014media}
Ralph, B.~C., Thomson, D.~R., Cheyne, J.~A., and Smilek, D. (2014).
\newblock Media multitasking and failures of attention in everyday life.
\newblock {\em Psychological research}, 78(5):661--669.

\bibitem[Ramzan et~al., 2019]{ramzan2019survey}
Ramzan, M., Khan, H.~U., Awan, S.~M., Ismail, A., Ilyas, M., and Mahmood, A.
  (2019).
\newblock A survey on state-of-the-art drowsiness detection techniques.
\newblock {\em IEEE Access}, 7:61904--61919.

\bibitem[Rasmussen, 1983]{rasmussen1983skills}
Rasmussen, J. (1983).
\newblock Skills, rules, and knowledge; signals, signs, and symbols, and other
  distinctions in human performance models.
\newblock {\em IEEE transactions on systems, man, and cybernetics},
  SMC-13(3):257--266.

\bibitem[Reason, 1990]{reason1990}
Reason, J. (1990).
\newblock {\em {Human Error}}.
\newblock Cambridge University Press.

\bibitem[Sahayadhas et~al., 2012]{sahayadhas2012detecting}
Sahayadhas, A., Sundaraj, K., and Murugappan, M. (2012).
\newblock {Detecting Driver Drowsiness Based on Sensors: A Review}.
\newblock {\em Sensors}, 12(12):16937--16953.

\bibitem[{Sikander} and {Anwar}, 2019]{Sikander2019}
{Sikander}, G. and {Anwar}, S. (2019).
\newblock {Driver Fatigue Detection Systems: A Review}.
\newblock {\em IEEE Transactions on Intelligent Transportation Systems},
  20(6):2339--2352.

\bibitem[Sun et~al., 2010]{sun2010activity}
Sun, F.-T., Kuo, C., Cheng, H.-T., Buthpitiya, S., Collins, P., and Griss, M.
  (2010).
\newblock {Activity-aware Mental Stress Detection Using Physiological Sensors}.
\newblock In {\em International conference on Mobile computing, applications,
  and services}, pages 282--301. Springer.

\bibitem[Sun et~al., 2018]{sun2018neural}
Sun, M., Tsujikawa, M., Onishi, Y., Ma, X., Nishino, A., and Hashimoto, S.
  (2018).
\newblock {A neural-network-based investigation of eye-related movements for
  accurate drowsiness estimation}.
\newblock In {\em International Conference of the IEEE Engineering in Medicine
  and Biology Society (EMBC)}, pages 5207--5210.

\bibitem[Swain and Guttmann, 1983]{swain1983handbook}
Swain, A.~D. and Guttmann, H.~E. (1983).
\newblock Handbook of human reliability analysis with emphasis on nuclear power
  plant applications.
\newblock {\em NUREG/CR-1278, SAND 80-0200}.

\bibitem[Tsujikawa et~al., 2018]{tsujikawal2018drowsiness}
Tsujikawa, M., Onishi, Y., Kiuchi, Y., Ogatsu, T., Nishino, A., and Hashimoto,
  S. (2018).
\newblock Drowsiness estimation from low-frame-rate facial videos using eyelid
  variability features.
\newblock In {\em International Conference of the IEEE Engineering in Medicine
  and Biology Society (EMBC)}, pages 5203--5206.

\bibitem[Uema and Inoue, 2017]{uema2017}
Uema, Y. and Inoue, K. (2017).
\newblock {JINS MEME Algorithm for Estimation and Tracking of Concentration of
  Users}.
\newblock In {\em Proceedings of the ACM International Joint Conference on
  Pervasive and Ubiquitous Computing and the ACM International Symposium on
  Wearable Computers (UbiComp/ISWC)}, pages 297--300.

\bibitem[Wang et~al., 2018]{wang2018novel}
Wang, H., Dragomir, A., Abbasi, N.~I., Li, J., Thakor, N.~V., and Bezerianos,
  A. (2018).
\newblock A novel real-time driving fatigue detection system based on wireless
  dry eeg.
\newblock {\em Cognitive neurodynamics}, 12(4):365--376.

\bibitem[Wijsman et~al., 2011]{wijsman2011towards}
Wijsman, J., Grundlehner, B., Liu, H., Hermens, H., and Penders, J. (2011).
\newblock Towards mental stress detection using wearable physiological sensors.
\newblock In {\em 2011 Annual International Conference of the IEEE Engineering
  in Medicine and Biology Society}, pages 1798--1801. IEEE.

\bibitem[Williamson et~al., 1993]{williamson1993human}
Williamson, J., Webb, R., Sellen, A., Runciman, W., and Van~der Walt, J.
  (1993).
\newblock Human failure: an analysis of 2000 incident reports.
\newblock {\em Anaesthesia and intensive care}, 21(5):678--683.

\bibitem[Yamada and Kobayashi, 2018]{yamada2018detecting}
Yamada, Y. and Kobayashi, M. (2018).
\newblock Detecting mental fatigue from eye-tracking data gathered while
  watching video: Evaluation in younger and older adults.
\newblock {\em Artificial intelligence in medicine}, 91:39--48.

\end{thebibliography}


\end{document}